\documentclass[fleqn,twoside]{article}
\usepackage{espcrc2}

\usepackage{graphicx}
\usepackage[figuresright]{rotating}
\input{epsf}
\newcommand{\beq}{\begin{equation}}
\newcommand{\eeq}{\end{equation}}
\newcommand{\bea}{\begin{eqnarray}}
\newcommand{\eea}{\end{eqnarray}}
\newcommand{\bmp}{\begin{minipage}}
\newcommand{\emp}{\end{minipage}}
\newcommand{\D}{\displaystyle}

\newcommand{\tr}{{\rm tr}}

\newcommand{\V}{{\cal V}}

\newcommand{\AmS}{{\protect\the\textfont2
  A\kern-.1667em\lower.5ex\hbox{M}\kern-.125emS}}

\title{RG Decimations and Confinement\thanks{Presented at {\it 
QCD Down Under}, Adelaide, Australia, March 10-19, 2004}}

\author{E. T. Tomboulis\address[MCSD]{Department of Physics and Astronomy, 
UCLA, \\
Los Angeles, CA 90095-1547, USA}%
        \thanks{Research partially supported by NSF-PHY-0309362.}}

\begin{document}

\begin{abstract}
We outline the steps in a derivation  
of the statement that the SU(2) gauge theory is in a 
confining phase for all values of the coupling, 
$0 < \beta <\infty$, defined at lattice 
spacing $a$. The approach employed is to obtain both 
upper and lower bounds for the partition 
function and the `twisted' partition function in terms of 
approximate decimation transformations. The behavior of the exact 
quantities is thus constrained by that of the easily computable 
bounding decimations. 
\vspace{1pc}
\end{abstract}

\maketitle

\section{Introduction} 

A very large body of work has been performed by the lattice 
community in recent years in an effort to isolate 
the types of configurations in the functional measure responsible 
for maintaining one confining phase for arbitrarily weak coupling 
in $SU(N)$ gauge theories. Thick vortices disordering the vacuum 
over long scales at negligible local free energy cost have emerged 
as a primary mechanism (see \cite{Gr} for a review and references). 
Nevertheless, a complete, direct derivation from first principles 
of this extraordinary and unique 
feature of $SU(N)$ theories (shared only by 
non-abelian ferromagnetic spin systems in $2$ dimensions)
has remained elusive for three decades. 

The difficulty stems from the multi-scale nature of the 
problem: passage from short distance ordered perturbative regime 
to long distance disordered non-perturbative confining regime. 
It can be addressed in principle only by a 
non-pertubative block-spinning procedure bridging short and 
long scales. Exact block-spinning schemes in gauge theories so 
far appear virtually intractable, both analytically and numerically. 

There are, however, approximate decimation procedures that 
can provide bounds on judicially chosen quantities. The idea is 
not new, but in the past only upper bounds were considered 
in this context. The basic strategy in the following is to 
obtain both upper and lower bounds for the 
partition function and the partition function in the presence of 
a `twist' (external center flux). The bounds are 
in terms of approximate 
decimations of the `potential moving' type, which can 
be explicitly computed to any accuracy. This leads to 
a rather simple construction constraining the behavior of the 
exact partition functions, and, through them, the exact vortex free energy 
and other order parameters, by that of the bounds. They thus are shown 
to exhibit 
confining behavior for all values of the inverse coupling,   
$0 < \beta < \infty$,  
defined at lattice spacing $a$ (UV cutoff) held fixed.  
Only the $SU(2)$ case is considered 
explicitly here, but the same development can be extended to 
general $SU(N)$.

\section{Decimations} 

We begin by some plaquette action, e.g the Wilson action 
$A_p(U,a) ={\beta\over 2}\;{\rm Re}\,\tr U_p$, defined 
at lattice spacing $a$.  
The character expansion of the exponential of the action: 
\beq 
F(U, a) =e^{A_p(U)} 
   = \sum_j\;F_j(\beta,a)\,d_j\,\chi_j(U) \label{exp}
\eeq
is given in terms of the Fourier coefficients: 
\beq F_j = \int\,dU\;F(U,a)\,{1\over d_j}\,\chi_j(U)\; .
\eeq
Here $\chi_j$ denotes the character of the $j$-th representation 
of dimension $d_j$.  Thus, for SU(2), 
$j=0, {1\over 2}, 1, {3\over 2}, \ldots$, and $d_j=(2j+1)$.
In terms of normalized coefficients: 
\beq 
c_j = {\D F_j \over \D F_0} \;,
\eeq
one then has  
\bea
F(U, a) &=&  F_0\,\Big[\, 1 + \sum_{j\not= 0} d_j\,c_j(\beta)\,\chi_j(U)\,
 \Big] \nonumber \\
    & \equiv & F_0\;f(U,a)  \label{nexp} 
\eea 
For a reflection positive action one necessarily has: 
\beq
F_j \geq 0 \;,\qquad \mbox{hence}\quad 1\geq c_j\geq 0\; , \qquad\quad 
\mbox{all}\quad j \;.
\eeq
The partition function on lattice $\Lambda$ is then  
\beq
Z_\Lambda(\beta) 
                    =F_0^{|\Lambda|}\; \int dU_\Lambda\;\prod_{p\in \Lambda}
\,f_p(U,a)
\;.\label{PF1}
\eeq

\begin{figure}[ht]
\centerline{\epsfxsize=3in\epsfbox{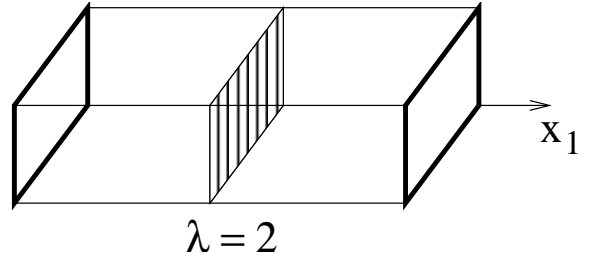}} 
\vspace{-0.2cm}  
\caption{Basic plaquette moving operation. \label{dec1}}
\vspace{-0.2cm} 
\end{figure}

We now consider RG decimation 
transformations $a \to \lambda a$. This involves partitioning the lattice 
in $d$-dimensional decimation cells of side length $\lambda a$.  
Simple approximate transformations of the `potential moving' type 
are implemented by  `weakening', i.e. decreasing the $c_j$'s  of 
plaquettes interior to the cells, 
and `strengthening', i.e. increasing $c_j$'s  of cell boundary plaquettes. 
The simplest scheme \cite{MK}, which is  
adopted in the following,  implements  
complete removal, $c_j=0$, of interior plaquettes. This may be 
pictured as moving the interior plaquette interactions to the 
cell boundaries. The operation can be decomposed into elementary 
moving steps along  each positive direction as illustrated 
for, say, the $x^1$-direction 
in Figure~\ref{dec1}. The $(\lambda -1 )$ interior plaquettes (shaded) 
are moved and  merged with the corresponding boundary plaquette (bold) into 
one boundary plaquette with renormalized interaction 
$A_p(U) \to \zeta A_p(U)$.

This basic operation is successively   
performed in all directions. Note that, in $d$ dimensions, there are 
$(d-2)$ normal directions into which a plaquette can move. A plaquette moved 
to the $(d-1)$-dimensional cell boundary can still be moved in 
$(d-3)$ directions inside the cell boundary. The moving operation terminates 
when all plaquettes have been moved in this manner to form a tiling 
of the $2$-dimensional faces of a lattice of spacing $\lambda a$. 
The integrations over the bonds of the tiling plaquettes inside 
each such face can now be performed exactly (being $2$-dimensional), 
thus merging the tiling plaquettes into one plaquette of the 
coarse lattice of spacing $\lambda a$. 

In terms of the definitions and notations introduced above, the end result 
can be concisely stated  as follows. 
Under successive decimations 
\bea 
& & a \to \lambda a \to \lambda^2 a \to \cdots \to \lambda^n a \nonumber \\
 & & \Lambda \to \Lambda^{(1)} \to \Lambda^{(2)} \to \cdots \to 
\Lambda^{(n)} \nonumber  
\eea
the resulting RG transformation rule is:
\beq
f(U,n-1) \to F_0(n)\,f(U,n) \label{RG1}
\eeq
with 
\beq
f(U,n)   =  \Big[ 1 + \sum_{j\not= 0} \,d_j
c_j(n)\,\chi_j(U) \Big] \;,\label{RG2}
\eeq
and 
\beq
c_j(n) = \hat{c}_j(n)^{\lambda^2}\; , \qquad 
F_0(n) = \hat{F}_0(n)^{\lambda^2}  \;,\label{RG3}
\eeq
where 
\beq
\hat{c}_j(n)\equiv \hat{F}_j(n)/\hat{F}_0(n) \leq 1  \;, \qquad j\not= 0\;,
\label{RG4}
\eeq
\beq
\hat{F}_j(n)= \int\,dU\;\Big[\,f(U,n-1)\,\Big]^\zeta\,
{1\over d_j}\,\chi_j(U) 
\; ,  \label{RG5}
\eeq
The renormalization parameter $\zeta$ controls by how much the 
plaquettes remaining after each decimation step have been strengthened 
to compensate for the removed plaquettes.  
What has been considered in the literature almost exclusively is 
the choice $\zeta=\lambda^{(d-2)}$. This  is essentially the original 
choice in \cite{MK}, and will be 
referred to as the MK choice. Here we 
consider $\zeta$ an adjustable parameter. 
 
The resulting partition function after $n$ decimation steps is:  
\bea 
Z_\Lambda(\beta, n) 
                    & = & \prod_{m=0}^n F_0(m)^{|\Lambda|/\lambda^{md}}
\nonumber \\ 
  & & \cdot \;\int dU_{\Lambda^{(n)}}\;\prod_{p\in\Lambda^{(n)}} \,f_p(U,n) 
\; .\label{PF2}
\eea

As a point of notation, the dependence of the quantities  
$F_0(n)$, $c_j(n)$, etc. on variables such as 
$\lambda$, $\zeta$, or  the set of couplings $\beta$, which characterize  
the choice of decimation and action at the original spacing $a$, will not 
be indicated explicitly unless specific reference to it is required.  

It is important to note that after each decimation step 
the resulting action retains the 
original {\it one-plaquette} form but will, in general,  
contain all representations:
\beq
A_p(U,n)= b_0(n) + \sum_{j\not=0}\; d_j\,\beta_j(n,\beta)\,\chi_j(U) \;.
\eeq 
Furthermore, among the effective couplings 
$\beta_j(n, \beta)$ some negative ones may in general occur. 
These features are present even after a single decimation 
step $a\to \lambda a$ starting with the usual single 
representation (fundamental) Wilson action.  

Preservation of the one-plaquette form of the action is of course what 
makes these decimations simple to explore. The rule specified by (\ref{RG1})-
(\ref {RG5}) is meaningful for any real (positive) $\zeta$. 
Here, however, a basic distinction can be made. 
For {\it integer} $\zeta$, 
the important property of positivity of the Fourier 
coefficients in (\ref{exp}), (\ref{nexp}):   
\beq
F_0(n)\geq 0 \quad , \qquad c_j(n)\geq 0 \;, \label{c+}
\eeq 
and hence reflection positivity are maintained at each decimation step. 
This, in general, is {\it not} the case for non-integer $\zeta$. 
Thus non-integer $\zeta$ results in approximate RG transformations that 
violate the reflection positivity of the theory (assuming a 
reflection positive starting action).\footnote{It is worth 
noting in this context that various numerical investigations 
of the standard MK recursions, at least for gauge theories, appear to 
have been carried out, for the most part, for fractional $\lambda$, 
($1< \lambda <2$), which corresponds to non-integer $\zeta$; e.g. 
see \cite{NT}.} From now on we assume that (\ref{c+}) holds. 

There are various other interesting features of such decimations. 
The following property, in particular, is  
important. One can show that, given the coefficients $c_j(n)$ 
after $n$ decimations, one has the general relation 
\beq
\sum_j \;c_j(n)\,\Big(\,\hat{c}_j(n+1, \zeta) - c_j(n)
\,\Big) \geq 0 \;.  
\label{cineq2}
\eeq
It follows from (\ref{cineq2}) that the ($l_2$) norm of 
the vector ${\bf \hat{c}}(n+1)$ formed from the $\hat{c}_j(n+1)$ 
coefficients is bigger  than that of the vector of the $c_j(n)$:
\beq 
||{\bf \hat{c}}(n+1)||_2 \geq ||{\bf c}(n)||_2 \; .\label{cineq3}
\eeq
All coefficients being positive, this implies  
that (\ref{cineq2}) must hold also component-wise, ie. 
\beq
\hat{c}_j(n+1) \geq c_j(n) \; ,\label{cineq4} 
\eeq
for at least a subset of components giving the dominant contribution 
to the norms.  For $\zeta=1$, it follows immediately from 
(\ref{RG4}) - (\ref{RG5}) that (\ref{cineq4}) holds as an equality 
for all $j$. For general $\zeta$, one finds by explicit 
numerical evaluations that, in fact,  (\ref{cineq4}) also 
holds for all $j$. This can be shown analytically  
in special cases and can probably be proved in general. 

Alternatively, instead of attempting such a proof, one may 
{\it impose} (\ref{cineq4}) as part of the specification  of the 
decimation transformation. This amounts to replacing (\ref{RG4}) by 
\bea
\hat{c}_j(n) & = & {\hat{F}_j(n)\over \hat{F}_0(n)}\;\Theta\Big[\,
{\hat{F}_j(n)\over \hat{F}_0(n)} - c_j(n-1)\,\Big] \nonumber \\
  & + & c_j(n-1)\;
\Theta\Big[\,c_j(n-1) - {\hat{F}_j(n)\over \hat{F}_0(n)}\,\Big] . 
\label{RG6}
\eea       
  
\section{The exact partition function}
Since our decimations are not exact decimation transformations, the partition 
function does not in general remain invariant under them. 
The subsequent development hinges on the following two basic 
propositions that can now be proved: 

{\bf (I)} With $\zeta=\lambda^{d-2}$:  
\beq
Z_\Lambda(\beta, n) \leq Z_\Lambda(\beta, n+1) \;.\label{I}
\eeq 

{\bf (II)} With $\zeta=1$: 
\beq
Z_\Lambda(\beta, n+1) \leq Z_\Lambda(\beta, n) \; .\label{II}
\eeq 
Note that for  $d=2$ (\ref{I}) - (\ref{II}) express the well-known 
fact that the decimations become exact. For $d>2$, in both (I), (II) one 
in fact has strict inequality. 

(I) says that modifying the couplings of the remaining plaquettes after 
decimation by taking $\zeta=\lambda^{d-2}$ (standard MK choice \cite{MK}) 
results into overcompensation (upper bound on the partition function). 
(II) says that decimating plaquettes while leaving the couplings of the 
remaining plaquettes unaffected results in a lower bound on the partition  
function. Translation invariance, convexity of the free energy 
and reflection positivity underlie (\ref{I}).

Consider now the, say, $(n-1)$-th decimation step  
with Fourier coefficients $c_j(n-1)$, which we relabel   
$c_j(n-1)=\tilde{c}_j(n-1)$.  
Given these $\tilde{c}_j(n-1)$, we proceed to compute the coefficients 
$F_0(n)$, $c_j(n)$ of the next decimation step 
according to (\ref{RG1})-(\ref{RG5}) above with $\zeta=\lambda^{d-2}$. 
 
Then introducing a parameter $\alpha$, ($0\leq \alpha$), define the 
{\it interpolating coefficients}:
\beq
\tilde{c}_j(n,\alpha)= \tilde{c}_j(n-1)^{\lambda^{2}(1-\alpha)}\,
 c_j(n)^\alpha \,, \label{inter1}
\eeq
so that  
\beq
\tilde{c}_j(n,\alpha)= \left\{ \begin{array}{lll}
c_j(n) & : &\alpha=1 \\ 
\tilde{c}_j(n-1)^{\lambda^{2}} & : & \alpha=0 
\end{array} \right. \label{inter2}
\eeq
The $\alpha=0$ value is that of the $n$-th step coefficients 
resulting from (\ref{RG3})-(\ref{RG5}) with $\zeta=1$.

Thus defining  the corresponding partition function
\bea    
Z_\Lambda(\beta, \alpha, n) &= & 
\big(\prod_{m=0}^{n-1} F_0(m)^{|\Lambda|/\lambda^{md}}\big)\;
F_0(n)^\alpha \nonumber \\ 
& & \cdot\:
\int dU_{\Lambda^{(n)}}\!\!\prod_{p \in \Lambda^{(n)}} 
f_p(U,n,\alpha) \label{PF3}
\eea
where 
\beq
f_p(U,n,\alpha)= \Big[ 1 + \sum_{j\not= 0} d_j\, \tilde{c}_j(n,\alpha)
\, \chi_j(U) \Big] \, ,
\eeq
we have from (\ref{I}), (\ref{II})), and (\ref{inter2}) above: 
\beq
Z_\Lambda(\beta, 0, n) \leq Z_\Lambda(\beta, n-1) 
\leq Z_\Lambda(\beta, 1, n)  \;. \label{III} 
\eeq
Now the partition function (\ref{PF3}) is a continuous 
in $\alpha$. So (\ref{III}) implies that, by continuity, 
there exist a value of $\alpha$: 
\[ \alpha=\alpha^{(n)}(\beta, \lambda, \Lambda)\,, \qquad  
0 < \alpha^{(n)}(\beta,\lambda, \Lambda) < 1  \]  
such that  
\beq 
Z_\Lambda(\beta, \alpha^{(n)}, n)= Z_\Lambda(\beta, n-1) \;.
\eeq
In other words there is an $\alpha$ at which the 
$n$-th decimation step partition function equals that obtained at 
the previous decimation step; the partition function does not 
change its value under the decimation step $\lambda^{n-1} a 
\to \lambda^n a$.

So starting at original spacing $a$, at every decimation step 
$m$, ($m=0,1,\cdots,n$), there exist a value $0< \alpha^{(m)} <1$ 
such that 
\beq
Z_\Lambda(\beta, \alpha^{(m+1)}, m+1)=
Z_\Lambda(\beta, \alpha^{(m)}, m)\, .\label{fixalpha}
\eeq 

This then gives, after $n$ successive decimations, a {\it  
representation} of the exact partition function in the form:
\bea    
Z_\Lambda(\beta) &= & 
F_0^{|\Lambda|}\; \int dU_\Lambda\;\prod_{p\in \Lambda}\,f_p(U,a) \nonumber \\
  & =& \prod_{m=0}^n F_0(m)^{\alpha^{(m)}|\Lambda|/\lambda^{md}}\;
\nonumber \\ 
& & \cdot\;
Z_{\Lambda^{(n)}}(\beta, n, \alpha^{(n)})
, \label{rep} 
\eea
where 
\bea
Z_{\Lambda^{(n)}}(\beta, n, \alpha^{(n)}) & \equiv & 
\int dU_{\Lambda^{(n)}}\;  \nonumber \\ 
& & \cdot\;\prod_{p\in \Lambda^{(n)}}\,f_p(U,n,\alpha^{(n)}) , 
\label{rep1} 
\eea
i.e. a representation in terms of the successive bulk free energy 
contributions from the $a \to \lambda \to \cdots \to \lambda^n a$ decimations 
and a one-plaquette effective action on the resulting 
lattice $\Lambda^{(n)}$.

At weak and strong coupling $\alpha^{(m)}$ may be estimated analytically. 
At large $\beta$, where the decimations approximate the free energy 
rather accurately, the appropriate $\alpha$ values are very close to 
unity. At strong coupling they may be estimated by comparison with 
the strong coupling expansion. On any finite lattice there is also a  
weak volume dependence as a correction which goes away    
as an inverse power of the lattice size. 

For most purposes the exact values of the $\alpha^{(m)}$'s, 
beyond the fact that are fixed between $0$ and $1$, are not 
immediately relevant. The main point of the representation (\ref{rep}) is 
that it can in principle relate the behavior of the exact theory to 
that of the easily computable approximate 
decimations.

Indeed, starting from the $\tilde{c}_j(n-1,\alpha^{(n-1)})$'s at the   
$(n-1)$-th  step, consider the coefficients at the next step, and compare   
those evaluated at $\alpha=\alpha^{(n)}$, i.e.   
\ $\tilde{c}_j(n,\alpha=\alpha^{(n)})$, to  
those evaluated at $\alpha=1$, i.e   
\ $\tilde{c}_j(n,\alpha=1)\equiv c_j(n)$. The latter  
will be referred to as 
the MK coefficients. (Recall that $\alpha=1 \Longleftrightarrow \zeta=
\lambda^{d-2}$, the standard MK choice. The absence 
of a tilde on a coefficient in the following 
always means that it is computed 
at $\alpha=1$.) According to (I), the MK coefficients give 
an upper bound. 

Now, (\ref{inter1}) implies that 
\beq 
\tilde{c}_j(n,\alpha) \leq 
(1-\alpha)\,\tilde{c}_j(n-1)^{\lambda^{2}} + 
\alpha \, c_j(n)\, , \label{inter3}
\eeq 
from which, by property (\ref{cineq4}), one has 
\beq
\tilde{c}_j(n,\alpha) \leq c_j(n) \qquad \mbox{for {\it any}} 
\quad 0\leq \alpha\leq 1 \;.\label{compineq}
\eeq 
This has the following important consequence.

(\ref{compineq}) says that the Fourier coefficients of the representation 
(\ref{rep}) are bounded from above by the MK coefficients 
($\alpha=1$). Thus, if  the $c_j(n)$'s are non-increasing, 
so are the $\tilde{c}_j(n,\alpha)$. 
The  $c_j(n)$'s must then approach a fixed point, and hence so 
must the $\tilde{c}_j(n,\alpha)$'s, since 
$c_j(n), \tilde{c}_j(n,\alpha)\geq 0$. 
Note the fact that this conclusion is independent of the specific 
value of the $\alpha$'s at every decimation step.

In particular, if the $c_j(n)$'s approach the strong coupling  
fixed point, i.e. $F_0\to 1$, $c_j(n) \to 0$ as $n\to \infty$, so must 
the $\tilde{c}_j(n,\alpha)$'s of the exact representation. 
{\it If the MK decimation coefficients flow to the confining regime, 
so do those in the exact representation (\ref{rep}).} 
As it is well-known by explicit numerical evaluation, the MK decimations 
for $SU(2)$ and $SU(3)$ indeed flow to the strong coupling confining regime   
for all $\beta<\infty$ and $d\leq 4$. Above the critical dimension 
$d=4$, the decimations result in free spin wave behavior. 

What do these results imply about the question of 
confinement in the exact theory? Though strongly suggestive, the 
fact that the long distance part, $Z_{\Lambda^{(n)}}$, in (\ref{rep}) flows in 
the confining regime does not suffice to answer the question. 
It is the combined contributions 
from all scales in (\ref{rep}) that combine to give one 
bulk quantity, the exact free energy 
$-\ln Z_\Lambda(\beta)$. As it is well-known, it is not possible 
to unambiguously determine the long distance behavior of the theory from 
that of a bulk quantity like the free energy. For that one needs to 
consider appropriate long distance order parameters.

\section{Order parameters - Vortex free energy}
The above derivation leading to the representation (\ref{rep}) for 
the partition function  cannot be applied in the presence of 
observables without modification. Thus, in the presence of 
operators involving external sources, such as the Wilson or 
't Hooft loop, translation invariance is lost. Reflection 
positivity is also reduced to hold only in the plane bisecting the loop. 
Fortunately, there are other order parameters that can  
characterize the possible phases of the theory while  
maintaining  translational invariance. They are the well-known vortex 
free energy, and its $Z(N)$ Fourier transform (electric flux free energy). 
They are in fact the natural order parameters in the present context 
since they are constructed out of partition functions. 
Recall that the vortex free energy is defined by 
\beq 
e^{-F_{v}(\tau)} = Z_\Lambda(\tau)/Z_\Lambda \; .\label{vfe} 
\eeq 
Here $Z_\Lambda(\tau)$ denotes the partition function  with action 
modified by the `twist' $\tau \in Z(N)$ for every plaquette on a 
coclosed set of plaquettes $\V$ winding through the 
periodic lattice in $(d-2)$ directions, i.e. through every 
$[\mu\nu]$-plane for fixed $\mu, \nu$:
\[ 
A_p(U_p) \to A_p(\tau U_p) \;, \qquad \mbox{if} \quad p\in \V\;. 
\]
One has $Z_\Lambda(1) = Z_\Lambda$. A nontrivial twist  ($\tau\not=1$)
represents a discontinuous 
gauge transformation on the set $\V$ which introduces 
$\pi_1(SU(N)/Z(N))$ vortex flux rendered topologically 
stable by being wrapped around the lattice torus.  
As indicated by the notation, 
$Z_\Lambda(\tau)$ depends only on the presence of the flux, and is 
invariant under changes in the exact location of $\V$. 
The vortex free energy is then 
the ratio of the partition function in the presence of this external 
flux to the partition function in the 
absence of the flux (the latter is what was considered above). 
As it is well-known the possible phases of a gauge theory (Higgs, 
Coulomb, confining) can be characterized by the behavior of 
(\ref{vfe}). Furthermore, 
by rigorous  correlation inequalities \cite{TY}, the Wilson loop, 
Wilson line correlators, and the 't Hooft loop can in turn be 
bounded by the vortex free energy and its $Z(N)$ Fourier transform. 
For $SU(2)$ the only nontrivial element is $\tau=-1$.  

The above development, in particular the derivation of (\ref{rep}), 
should then be repeated for $Z_\Lambda(-1)\equiv Z_\Lambda^{(-)}$. 
There is, however, a technical complication in obtaining the analog 
to (\ref{rep}) for $Z_\Lambda^{(-)}$. The presence of the flux  
reduces reflection positivity to hold only in planes 
perpendicular to the  directions in which 
the flux winds through the lattice. This may be circumvented by 
considering $Z_\Lambda + Z_\Lambda^{(-)}$ instead of $Z_\Lambda^{(-)}$. 
Indeed, for $Z_\Lambda + Z_\Lambda^{(-1)}$ reflection positivity is 
easily checked to again hold in all planes. The proof of propositions 
(I) and (II), and the subsequent derivation 
leading to (\ref{fixalpha}), and (\ref{rep})  can then be carried 
through for the quantity 
$Z_\Lambda + Z_\Lambda^{(-)}$ to obtain 
\[    
Z_\Lambda(\beta) + Z_\Lambda^{(-)}(\beta) = 
\prod_{m=0}^n F_0(m)^{\alpha^{(m)}|\Lambda|/\lambda^{md}} \]
\beq
 \qquad  \cdot \; \Big(\,Z_{\Lambda^{(n)}}(\beta, n, \alpha^{(n)}) + 
Z^{(-)}_{\Lambda^{(n)}}(\beta, n, \alpha^{(n)})\,\Big)\label{rep+}
\eeq
where 
\[ 
Z^{(-)}_{\Lambda^{(n)}} = \int dU_{\Lambda^{(n)}} 
\prod_{{p\notin \V \atop p\in \Lambda^{(n)}}}\,f_p(U,n,\alpha^{(n)}) \] 
\beq 
\qquad  \qquad \cdot \; 
\prod_{{p\in \V \atop p\in \Lambda^{(n)}}} \,f_p((-1)U,
n,\alpha^{(n)})\,\big). 
\label{rep+1} 
\eeq  
The values $\{\alpha^{(m)}\}$ in (\ref{rep+}), as fixed by 
the interpolating argument between the upper and lower bounds 
for the quantity $Z_\Lambda + Z_\Lambda^{(-)}$, are 
a priori distinct from those in (\ref{rep}) fixed by the analogous 
argument  for the quantity $Z_\Lambda$. It is not hard to see, however,  
that in fact they have to coincide for large lattices.  

Using (\ref{rep}) and (\ref{rep+}) in (\ref{vfe}) then gives 
for the vortex free energy:
\beq
e^{-F_{v}(\tau)}= { Z^{(-)}_{\Lambda^{(n)}}(\beta, n, \alpha^{(n)}) 
\over  Z_{\Lambda^{(n)}}(\beta, n, \alpha^{(n)})} \;.\label{vfe1}
\eeq 
It manifestly couples only to the 
long distance part as it should. 
Bulk (local) free energy contributions resulting from each 
successive decimation are unaffected by the presence of the flux 
and cancel. By our previous considerations, the
strategycoefficients $\tilde{c}_j(n,\alpha^{(n)})$ occurring in 
$Z_{\Lambda^{(n)}}$, $Z^{(-)}_{\Lambda^{(n)}}$ in (\ref{vfe1}) 
are bounded (regardless of the specific values 
of the $\alpha^{(n)}$ parameters) by the corresponding MK 
coefficients $c_j(n)\equiv \tilde{c}_j(n,\alpha=1)$. By 
taking $n$ large enough, the latter 
tend to zero no matter how large the initial $\beta$ is chosen.  
Thus taking $n$ large enough to enter the strong coupling regime, 
(\ref{vfe1}) can be evaluated exactly within the convergent 
strong coupling cluster expansion in terms of the  
$\tilde{c}_j(n,\alpha^{(n)})$'s by standard computations. 
The vortex free energy 
is thus explicitly shown to exhibit confining behavior. This,  
by the known inequalities \cite{TY} relating the vortex free energy 
to the Wilson loop, then also necessarily implies area law for 
the latter. 

An expanded account with the proofs of (I), (II) and various other 
statements above will appear elsewhere. \\
\\
 
\vspace{0.05cm} 
I am grateful to the organizers of the QCD Down Under Workshop 
for organizing such a stimulating meeting in a thoroughly enjoyable 
environment in Adelaide and the Barossa Valley, and to the 
participants for a great many discussions.

\end{document}